\documentclass[11pt,a4paper]{article}

\def\fsd{{\ \succeq_{1}\ }}
\def\ssd{{\ \succeq_{2}\ }}
\def\trieq{\triangleq}

\usepackage[latin1]{inputenc}
\usepackage{tikz}
\usetikzlibrary{trees}
\usepackage{verbatim}
\usepackage{amsfonts}
\usepackage{amssymb}
\usepackage{amsmath}
\usepackage{amsthm}
\usepackage{dsfont}
\usepackage{bbm}
\usepackage{graphics}
\usepackage{graphicx}
\usepackage{epsfig}
\usepackage{mathrsfs}
\usepackage[all]{xy}
\usepackage{color}

\newtheorem{theorem}{Theorem}[section]

\newtheorem{corollary}[theorem]{Corollary}
\newtheorem{definition}[theorem]{Definition}

\newtheorem{lemma}[theorem]{Lemma}
\newtheorem{proposition}[theorem]{Proposition}
\newtheorem{remark}[theorem]{Remark}

\numberwithin{equation}{section}

\def\11{\mathbbold{1}}
\def\qed{{\hbox{ }\hfill {$\Box$}}}

\def\proof{\noindent\textbf{Proof. }}

\def\VaR{\mathrm{VaR}}

\def\AVaR{\mathrm{AVaR}}

\DeclareMathOperator*{\esssup}{ess\,sup\,}

\def\Pbb{\mathbb{P}}

\def\Rbb{\mathbb{R}}

\def\Fc{\mathcal{F}}
\def\Gc{\mathcal{G}}

\def\Qc{\mathcal{Q}}

\def\Wc{\mathcal{W}}

\def\Asc{\mathscr{A}}
\def\Bsc{\mathscr{B}}
\def\Csc{\mathscr{C}}
\def\Dsc{\mathscr{D}}

\def\Msc{\mathscr{M}}

\def\Qsc{\mathscr{Q}}

\title{Monetary Risk Measures\thanks{Supported by National Key R\&D Program of China (NO. 2018YFA0703900) and the Major Project of National Social Science Foundation of China (NO.19ZDA091): Research on Dynamic Monitoring of Local Financial Operation and Early Warning of Systemic Risk.}}
\author{%(Comments Welcome)\\  \\
Guangyan Jia\thanks{Zhongtai Securities Institute for Financial Studies, Shandong University, Jinan 250100, China; Email: jiagy@sdu.edu.cn.}
\and
Jianming Xia\thanks{RCSDS, NCMIS, Academy of Mathematics and Systems Science,
Chinese Academy of Sciences, Beijing 100190, China; Email:
xia@amss.ac.cn.}
\and
Rongjie Zhao\thanks{School of Mathematical Sciences, University of Chinese Academy of Sciences, Beijing 100049, China; Email: zrj2010@hotmail.com.}
}

\date{}%{November 12, 2020}

\begin{document}

\maketitle

\begin{abstract}
In this paper, we study general monetary risk measures (without any
convexity or weak convexity). A monetary (respectively, positively
homogeneous) risk measure can be characterized as the lower envelope
of a family of convex (respectively,  coherent) risk measures. The
proof does not depend on but easily leads to the classical
representation theorems for convex and coherent risk measures. When
the law-invariance and the SSD (second-order stochastic
dominance)-consistency are involved, it is not the convexity
(respectively, coherence) but the comonotonic convexity
(respectively, comonotonic coherence) of risk measures that can be
used for such kind of lower envelope characterizations in a unified
form. The representation of a law-invariant risk measure in terms of
VaR is provided.  
%As a case study, we also investigate the risk
%measures based on rank-dependent utilities.

\end{abstract}

\section{Introduction}

It is an important subject to measure the risk of a financial position.
VaR (value at risk) has long been a standard risk measure in industry,
whether by choice or by regulation. However, VaR has been criticized
in both academia and industry, mainly for two shortcomings: (1) VaR
focuses on the probability of loss, regardless of the magnitude, and
therefore, fails to capture ``tail risk"; (2) VaR is not sub-additive, and therefore, violates the principle of diversification.

Recognizing the shortcomings of VaR, in their seminal work, Artzner, Delbaen, Eber and Heath (1999) argued that a good risk measure
should satisfy a set of reasonable axioms (monotonicity, translation-invariance, sub-additivity and positive homogeneity),
leading to the so-called coherent risk measures.\footnote{For representations of coherent risk measures on general probability spaces,
see Delbaen (2002).}
%One of the most popular coherent risk measures is AVaR (average value at
%risk\footnote{Also sometimes termed ``expected shortfall" or
%``conditional value at risk" in the literature.}).
%was suggested by both
%academia and industry; see, e.g., Rockafellar and Uryasev (2000,
%2002), Acerbi and Tasche (2002), Alexander and Baptista (2006),
%Embrechts et al. (2014), and BCBS (2016).
F\"ollmer and Schied (2002), as well as Frittelli and Rosazza Gianin
(2002),  argued that, in many situations, the risk of a position
might increase in a nonlinear way with the size of the position and
suggested to relax the sub-additivity and positive homogeneity to
the convexity, leading to the so-called convex risk measures. In
addition, Song and Yan (2006, 2009) and Kou, Peng and Heyde (2013)
further argued that the sub-additivity and the convexity for all
risks might be too restrictive and suggested relaxations to the
comonotonic sub-additivity or the comonotonic convexity, leading to
the so-called comonotonic sub-additive or comonotonic convex risk
measures.\footnote{Some other relaxations, from the
translation-invariance and convexity to the quasi-convexity (and
cash-subadditivity), were carried out by El Karoui and Ravanelli
(2009) and Cerreia-Vioglio et al. (2011).} For discussions on
law-invariance of risk measures, see among others Kusuoka (2001),
Frittelli and Rosazza Gianin (2005), Jouini, Schachermayer and Touzi
(2006), and Song and Yan (2009).

All kinds of risk measures above satisfy two basic axioms:
monotonicity and translation-invariance. A risk measure satisfying
the two basic axioms is usually called a monetary risk measure; see,
e.g., F\"ollmer and Schied (2016). In this paper, our focus is on
the characterizations and representations of a general monetary risk
measure (without any convexity or weak convexity).

A closely related work to ours is Mao and Wang (2020), which is
fundamental and inspiring. They argued that a risk measure should be
consistent with risk aversion, which can be described by the
consistency with respect to SSD (second-order stochastic dominance),
leading to the so-called SSD-consistent risk measures. They provided
characterizations and representations, as well as some applications,
of SSD-consistent risk measures. Although no convexity is imposed a
priori on an SSD-consistent risk measure, we will see that an
SSD-consistent risk measure still satisfies a kind of weak
convexity, which is called ID-convexity (identical-distribution
convexity) in this paper; see Theorem \ref{thm:idcv} below.

The most famous non convex risk measure is VaR, which is excluded by Mao and Wang (2020) since it is not SSD-consistent.
Every risk measure has its advantages as well as disadvantages. No one can do everything better than another. We do not argue for one risk measure against another, but investigate general monetary risk measures, which include VaR as a special case.

%Although VaR has been criticized
%in both academia and industry, it still has advantages. More recently, Kou et al. (2013) and Cont et al. (2010, 2013) noted that VaR leads to more robust estimates than coherent risk measures when using historic data to estimate the risk of a portfolio.

The contribution and the structure of our paper are as follows (the results for positively homogeneous risk measures are similar and hence not list here for the brevity).

Section \ref{sec:rm:as} collects some preliminaries about risk measures and acceptance sets.

In Section \ref{sec:mrm}, we show that a monetary risk measure $\rho$ is the lower envelope of a family of convex risk measures and has the following representation
$$\rho(X)=\min_{\lambda\in\Lambda}\sup_{Q\in\Msc_1(P)}\left(E_Q[-X]-\alpha_\lambda(Q)\right),$$
where $\Msc_1(P)$ denotes all probability measures which are absolutely continuous with respect to $P$ and $\{\alpha_\lambda\mid\lambda\in\Lambda\}$ is a family of convex functionals $\alpha_\lambda: \Msc_1(P)\to(-\infty,\infty]$.
As a consequence, we have, for every monetary (similarly, positively homogeneous) risk measure $\rho$,
$$\rho(X)=\inf\{h(X)\mid h \mbox{ convex (coherent) and }h\ge\rho\},$$
which extends the following equality in Artzner, Delbaen, Eber and Heath (1999, Proposition 5.2):
\begin{align}\label{eq:var:cohe}
\VaR_t(X)=\inf\{h(X)\mid h \mbox{ coherent and }h\ge\VaR_t\}.
\end{align}
It is interesting that the proof of the representation theorem for general monetary risk measures does not depend on but easily leads to the classical representation theorem for convex and coherent risk measures, as Corollary \ref{cor:convex} and Remark \ref{rmk:phrm:rep} show.

In Section \ref{sec:LI}, we show that a law-invariant monetary risk measure has the following representation
\begin{align*}
\rho(X)=\min_{\lambda\in \Lambda}\sup_{g\in\Gc}\left(\int_0^1\VaR_t(X)g(t)dt-\alpha_\lambda(g)\right),
\end{align*}
where $\Gc$ denotes all probability density functions on $(0,1)$ and
 $\{\alpha_\lambda\mid\lambda\in\Lambda\}$ is a family of convex functionals $\alpha_\lambda: \Gc\to(-\infty,\infty]$.

In Section \ref{sec:ssd}, we investigate SSD-consistent risk measures. Some results of Mao and Wang (2020) are recovered and some other representations of SSD-consistent risk measures are provided. We provide another equivalent formulation of SSD-consistency: SSD-consistency is equivalent to the combination of law-invariance, the Fatou property and ID-convexity.  As far as we know, the ID-convexity has not yet been introduced in the literature to study risk measures.

From Sections  \ref{sec:mrm}--\ref{sec:ssd}, we know each monetary risk measure is the lower envelope of a family of convex risk measures. When the law-invariance is involved, it is a ``natural" expectation that each law-invariant monetary risk measure is the lower envelope of a family of law-invariant convex risk measures.  This ``natural" expectation, however, is not correct since law-invariant convex risk measures are SSD-consistent and hence so is the lower envelope of some of them. For example,
\begin{align}\label{ineq:var:cohe:L}
\VaR_t(X)\ne\inf\{h(X)\mid h \mbox{ law-invariant and coherent and }h\ge\VaR_t\}.
\end{align}
A natural question arises:
\begin{quote}
\textit{What is the ``good" property such that general (respectively, law-invariant, SSD-consistent) monetary risk measures can be represented in a unified form: they are the lower envelopes of a family of general (respectively, law-invariant, SSD-consistent) monetary risk measures having the ``good" property?}
\end{quote}
Section \ref{sec:unif} replies to this question. It turns out that the comonotonic convexity is the desired ``good" property.

%In Section \ref{sec:rdu}, we study the risk measures based on RDU (rank-dependent utilities), which include as special cases those based on expected utilities.

\section{Preliminaries: Risk Measures and Acceptance Sets}\label{sec:rm:as}

Let $(\Omega,\Fc,P)$ be a complete probability space. We use $L^\infty(P)$ to denote $L^\infty(\Omega,\Fc,P)$ for brevity. We study risk measures on $L^\infty(P)$. Some basic definitions and facts are recalled in this section.

\begin{definition}
A mapping $\rho: L^\infty(P)\to\Rbb$ is called a monetary risk measure if it satisfies the following two conditions for all $X,Y\in L^\infty(P)$.
\begin{itemize}
  \item Monotonicity: If $X\le Y$ $P$-a.s., then $\rho(X)\ge\rho(Y)$.
  \item Translation-Invariance: If $m\in\Rbb$, then $\rho(X+m)=\rho(X)-m$.
\end{itemize}
\end{definition}

\begin{definition}
A monetary risk measure $\rho: L^\infty(P)\to\Rbb$ is called a positively homogeneous risk measure if it satisfies
\begin{itemize}
  \item Positive Homogeneity: $\rho(\alpha X)=\alpha\rho(X)$ for $X\in L^\infty(P)$ and $\alpha\ge0$.
\end{itemize}
\end{definition}

\begin{definition}
A monetary risk measure $\rho: L^\infty(P)\to\Rbb$ is called a convex risk measure if it satisfies
\begin{itemize}
  \item Convexity: $\rho(\alpha X+(1-\alpha)Y)\le\alpha\rho(X)+(1-\alpha)\rho(Y)$ for $X,Y\in L^\infty(P)$ and $\alpha\in[0,1]$.
\end{itemize}
\end{definition}

\begin{definition}
A monetary risk measure $\rho: L^\infty(P)\to\Rbb$ is called a coherent risk measure if it is convex and positively homogeneous.
\end{definition}

Given a monetary risk measure $\rho$, its \textit{acceptance set} $\Asc_\rho$ is given by
$$\Asc_\rho\trieq\{X\in L^\infty(P)\mid\rho(X)\le0\}.$$
Given a subset $\Asc\subseteq L^\infty(P)$, let the mapping $\rho_\Asc$ be given by
$$\rho_\Asc(X)\trieq\inf\{m\in\Rbb\mid X+m\in\Asc\}, \quad X\in L^\infty(P).$$
The following proposition summarizes the relation between monetary
risk measures and their acceptance sets; see, e.g., F\"ollmer and
Schied (2016, Propositions 4.6--4.7).

\begin{proposition}\label{prop:rm:set}
A mapping $\rho: L^\infty(P)\to\Rbb$ is a monetary risk measure if and only if $\rho=\rho_\Asc$ for a nonempty subset $\Asc\subseteq L^\infty(P)$ satisfying the following two conditions.
\begin{description}
\item[(i)] $\inf\{m\in\Rbb\mid m\in\Asc\}>-\infty$;
\item[(ii)] $X\in\Asc$, $Y\in L^\infty(P)$, $Y\ge X$ $P$-a.s.\, $\Rightarrow$\, $Y\in\Asc$.
\end{description}
If it is the case, the set $\Asc$ can be chosen as the acceptance set $\Asc_\rho$ of $\rho$. Moreover, we have
\begin{itemize}
\item $\rho$ is a positively homogeneous  risk measure if and only if $\Asc_\rho$ is a cone;
\item $\rho$ is a convex risk measure if and only if $\Asc_\rho$ is a convex set;
\item $\rho$ is a coherent risk measure if and only if $\Asc_\rho$ is a convex cone.
\end{itemize}
\end{proposition}

The following easy lemma will be frequently used, whose proof is omitted.

\begin{lemma}\label{lma:inf}
Assume $\{\Asc_\lambda\mid\lambda\in\Lambda\}$ is a class of subsets of $L^\infty(P)$ and each $\Asc_\lambda$ satisfies conditions (i)--(ii) of Proposition \ref{prop:rm:set}. Let $\Asc=\bigcup_{\lambda\in\Lambda}\Asc_\lambda$.
Then $\rho_\Asc(X)=\inf_{\lambda\in\Lambda}\rho_{\Asc_\lambda}(X)$ for $X\in L^\infty(P)$.
\end{lemma}

\section{Representations of Monetary Risk Measures}\label{sec:mrm}

Let $$\Msc_1(P)=\Msc_1(\Omega,\Fc,P)$$ denote the set of all probability measures on $(\Omega,\Fc)$ which are absolutely continuous with respect to $P$.  Let $$\Msc_{1,f}(P)=\Msc_{1,f}(\Omega,\Fc,P)$$ denote the set of all \textit{finitely additive} probability measures on $(\Omega,\Fc)$ which are absolutely continuous with respect to $P$. It is well known that $\Msc_{1,f}(P)$ is compact under the weak*-topology, which is induced by $L^\infty(P)$.

\medskip

The following theorem shows that a monetary risk measure is the lower envelope of a family of convex risk measures. % having the Fatou property.

\begin{theorem}\label{thm:mrm:rep:P}
For a mapping $\rho: L^\infty(P)\to\Rbb$, the following assertions are equivalent.
\begin{description}
\item[(a)] $\rho$ is a monetary risk measure.
\item[(b)] There exists a family $\{\alpha_\lambda\mid\lambda\in\Lambda\}$ of convex functionals $\alpha_\lambda: \Msc_1(P)\to(-\infty,\infty]$ such that
$$\rho(X)=\min_{\lambda\in\Lambda}\sup_{Q\in\Msc_1(P)}\left(E_Q[-X]-\alpha_\lambda(Q)\right)\quad\mbox{ for }X\in L^\infty(P).$$
\item[(c)] There exists a family $\{\alpha_\lambda\mid\lambda\in\Lambda\}$ of lower semi-continuous (under the weak* topology), convex functionals $\alpha_\lambda: \Msc_{1,f}(P)\to(-\infty,\infty]$ such that
$$\rho(X)=\min_{\lambda\in\Lambda}\max_{Q\in\Msc_{1,f}(P)}\left(E_Q[-X]-\alpha_\lambda(Q)\right)\quad\mbox{ for }X\in L^\infty(P).$$
\item[(d)] There exists a family $\{\rho_\lambda\mid\lambda\in\Lambda\}$ of convex risk measures on $L^\infty(P)$ such that
%each $\rho_\lambda$ has the Fatou property and
$$\rho(X)=\min_{\lambda\in\Lambda}\rho_\lambda(X)\quad\mbox{ for }X\in L^\infty(P).$$
\item[(e)] For each $X\in L^\infty(P)$,
\begin{align}\label{eq:rho:h:1}
\rho(X)=\inf\{h(X)\mid h \mbox{ is a convex risk measure and }h\ge\rho\}.
\end{align}
\end{description}
\end{theorem}

\proof We only prove ``(a)$\Rightarrow$(b)" and ``(a)$\Rightarrow$(c)", since ``(b)$\Rightarrow$(d)$\Rightarrow$(e)$\Rightarrow$(a)"  and ``(c)$\Rightarrow$(d)" are obvious.

\begin{description}
\item[(a)$\Rightarrow$(b):] Assume $\rho$ is a monetary risk measure. For any
$Z\in\Asc_\rho$, let
$$\Asc(Z)=\{Y\in L^\infty(P)\mid Y\ge Z\ P\mbox{-a.s.}\}.$$
Firstly, each $\Asc(Z)$ is obviously a convex subset of $L^\infty(P)$ satisfying conditions (i)--(ii) of Proposition \ref{prop:rm:set}. Then by Proposition \ref{prop:rm:set}, each $\rho_{\Asc(Z)}$ is a convex risk measure.
Next, $\Asc_\rho=\bigcup_{Z\in\Asc_\rho}\Asc(Z)$. Then by Lemma \ref{lma:inf},
\begin{align}\label{eq:rhoZ}
\rho(X)=\rho_{\Asc_\rho}(X)=\inf_{Z\in\Asc_\rho}\rho_{\Asc(Z)}(X),\quad \forall\, X\in L^\infty(P).
\end{align}
Moreover, for each $X\in L^\infty(P)$, we have $Z_0=X+\rho(X)\in\Asc_\rho$ and
$$\rho_{\Asc(Z_0)}(X)=\inf\{m\in\Rbb\mid X+m\ge  X+\rho(X)\}=\rho(X),$$
which implies that the infimum in \eqref{eq:rhoZ} can be attained and hence
$$\rho(X)=\min_{Z\in\Asc_\rho}\rho_{\Asc(Z)}(X).$$
Finally, for each $X\in L^\infty(P)$ and $Z\in\Asc_\rho$, we have
\begin{align}\label{eq:rho:eqzx}
\begin{aligned}
\rho_{\Asc(Z)}(X)%=&\inf\{m\in\Rbb\mid X+m\in\Asc(Z)\}\\
=&\inf\{m\in\Rbb\mid X+m\ge Z\ P\mbox{-a.s.}\}\\
%=&\inf\{m\in\Rbb\mid m\ge Z-X\ P\mbox{-a.s.}\}\\
=&\,\esssup(Z-X)\\
=&\sup_{Q\in\Msc_1(P)}E_Q[Z-X]\\
=&\sup_{Q\in\Msc_1(P)}(E_Q[-X]-\alpha_Z(Q)),
\end{aligned}
\end{align}
where $\alpha_Z(Q)=E_Q[-Z]$ for each $Q\in\Msc_1(P)$.
Therefore, $\{\alpha_Z\mid Z\in\Asc_\rho\}$ is a  desired family of convex functionals on $\Msc_1(P)$.

\item[(a)$\Rightarrow$(c):] According the proof of ``(a)$\Rightarrow$(b)", we have
\begin{align*}
\rho_{\Asc(Z)}(X)=&\,\esssup(Z-X)\\
=&\max_{Q\in\Msc_{1,f}(P)}E_Q[Z-X]\\
=&\max_{Q\in\Msc_{1,f}(P)}(E_Q[-X]-\alpha_Z(Q)),
\end{align*}
where $\alpha_Z(Q)=E_Q[-Z]$ for each $Q\in\Msc_{1,f}(P)$.
Therefore, $\{\alpha_Z\mid Z\in\Asc_\rho\}$ is a  desired family of convex functionals on $\Msc_{1,f}(P)$.\qed
\end{description}

%\item[(b)$\Rightarrow$(c):]  By \eqref{eq:rho:eqzx}, each $\rho_{\Asc(Z)}$ has the Fatou property. Therefore, $\{\rho_{\Asc(Z)}\mid Z\in\Asc_\rho\}$ is a desired family of convex risk measures.
%
%\item[(c)$\Rightarrow$(a):] It is obvious.\qed
%\end{description}

\begin{remark}\label{rmk:mrm}
\begin{description}
\item[(a)]
From the proof of Theorem \ref{thm:mrm:rep:P}, where a similar argument of Mao and Wang (2020) is used, we can see that a monetary risk measure $\rho: L^\infty(P)\to\Rbb$ has the following representations
\begin{align*}
\rho(X)=&\min_{Z\in\Asc_\rho}\sup_{Q\in\Msc_1(P)}(E_Q[-X]-E_Q[-Z])\\
=&\min_{Z\in\Asc_\rho}\max_{Q\in\Msc_{1,f}(P)}(E_Q[-X]-E_Q[-Z]),\quad\forall\,X\in L^\infty(P).
\end{align*}
\item[(b)]
The characterizations \eqref{eq:rho:h:1} and \eqref{eq:rho:h:2}
below are extensions of equality \eqref{eq:var:cohe}, which is from
Artzner, Delbaen, Eber and Heath (1999).
\item[(c)] For some related applications in decision theory under uncertainty, see Xia (2020).
\end{description}
\end{remark}

As the next corollary shows, the proof of Theorem \ref{thm:mrm:rep:P} does not depend on but leads to the classical representation theorem for convex risk measures of F\"ollmer and Schied (2002) and Frittelli and Rosazza Gianin (2002).

\begin{corollary}\label{cor:convex}
Assume $\rho: L^\infty(P)\to\Rbb$ is a convex risk measure. Then there exists a  lower-semicontinuous convex functional $\alpha: \Msc_{1,f}(P)\to(-\infty,\infty]$ such that
$$\rho(X)=\max_{Q\in\Msc_{1,f}(P)}(E_Q[-X]-\alpha(Q))\quad\mbox{ for }X\in L^\infty(P).
$$
\end{corollary}

\proof The convexity of $\rho$ implies that its acceptance set $\Asc_\rho$ is convex. Moreover, $\Msc_{1,f}(P)$ is a weak*-compact and convex set. From
the representations in Remark \ref{rmk:mrm} and by the minmax theorem\footnote{The minmax theorem is frequently used in this paper, for which see, e.g., Mertens, Sorin and Zamir (2015, Theorem I.1.1).}, we have
for each $X\in L^\infty(P)$ that
\begin{align*}
\rho(X)=&\min_{Z\in\Asc_\rho}\max_{Q\in\Msc_{1,f}(P)}(E_Q[-X]-E_Q[-Z])\\
=&\max_{Q\in\Msc_{1,f}(P)}\inf_{Z\in\Asc_\rho}(E_Q[-X]-E_Q[-Z])\\
=&\max_{Q\in\Msc_{1,f}(P)}(E_Q[-X]-\alpha(Q)),
\end{align*}
where $\alpha(Q)=\sup_{Z\in\Asc_\rho}E_Q[-Z]$ defines the desired lower-semicontinuous convex functional on $\Msc_{1,f}(P)$.\qed

\medskip

The following theorem shows that a positively homogeneous risk measure is the lower envelope of a family of coherent risk measures.
% having the Fatou property.

\begin{theorem}\label{thm:phrm:rep:P}
For a mapping $\rho: L^\infty(P)\to\Rbb$, the following assertions are equivalent.
\begin{description}
\item[(a)] $\rho$ is a positively homogeneous risk measure.
\item[(b)] There exists a family $\{\Qsc_\lambda\mid\lambda\in\Lambda\}$ of nonempty, weak*-compact, convex subsets of $\Msc_{1,f}(P)$ such that
$$\rho(X)=\min_{\lambda\in\Lambda}\max_{Q\in\Qsc_\lambda}E_Q[-X]\quad\mbox{ for }X\in L^\infty(P).$$
\item[(c)] There exists a family $\{\rho_\lambda\mid\lambda\in\Lambda\}$ of coherent risk measures on $L^\infty(P)$ such that
%each $\rho_\lambda$ has the Fatou property and
$$\rho(X)=\min_{\lambda\in\Lambda}\rho_\lambda(X)\quad\mbox{ for }X\in L^\infty(P).$$
\item[(d)] For each $X\in L^\infty(P)$,
\begin{align}\label{eq:rho:h:2}
\rho(X)=\inf\{h(X)\mid h \mbox{ is a coherent risk measure and }h\ge\rho\}.
\end{align}
\end{description}
\end{theorem}

\proof We only prove ``(a)$\Rightarrow$(b)", since ``(b)$\Rightarrow$(c)$\Rightarrow$(d)$\Rightarrow$(a)" is obvious.

Assume $\rho$ is a positively homogeneous risk measure. For any
$Z\in\Asc_\rho$, let $\Asc(Z)$ be given as in the proof of Theorem \ref{thm:mrm:rep:P}.
By Proposition \ref{prop:rm:set}, we know that $\Asc_\rho$ is a cone, which combined with Remark \ref{rmk:mrm} imply that, for each $X\in L^\infty(P)$,
\begin{align*}
\rho(X)=&\min_{Z\in\Asc_\rho}\,\max_{Q\in\Msc_{1,f}(P)}(E_Q[-X]+E_Q[Z])\\
=&\min_{Z\in\Asc_\rho,\, \alpha\ge0}\,\max_{Q\in\Msc_{1,f}(P)}(E_Q[-X]+E_Q[\alpha Z])\\
=&\min_{Z\in\Asc_\rho}\inf_{\alpha\ge0}\,\max_{Q\in\Msc_{1,f}(P)}(E_Q[-X]+E_Q[\alpha Z])\\
=&\min_{Z\in\Asc_\rho}\,\max_{Q\in\Msc_{1,f}(P)}\inf_{\alpha\ge0}(E_Q[-X]+E_Q[\alpha Z])\quad\mbox{(by the minmax theorem)}\\
=&\min_{Z\in\Asc_\rho}\,\max_{Q\in\Msc_{1,f}(P)}\left(E_Q[-X]+\inf_{\alpha\ge0}E_Q[tZ]\right)\\
=&\min_{Z\in\Asc_\rho}\,\max_{\substack{Q\in\Msc_{1,f}(P)\\ E_Q[Z]\ge0}}E_Q[-X],
\end{align*}
where we can apply the minmax theorem
%\footnote{See, e.g., Mertens, Sorin and Zamir (2015, Theorem I.1.1).}
since $\Msc_{1,f}(P)$ is weak*-compact.
Let $\Qsc_Z=\{Q\in\Msc_{1,f}(P)\mid E_Q[Z]\ge0\}$ for each $Z\in\Asc_\rho$. Then
$\{\Qsc_Z\mid Z\in\Asc_\rho\}$ is a desired family of subsets of $\Msc_{1,f}(P)$.
\qed

\begin{remark}\label{rmk:phrm:rep} From the proof of Theorem \ref{thm:phrm:rep:P}, we can see that a positively homogeneous risk measure $\rho: L^\infty(P)\to\Rbb$ has the following representation
$$\rho(X)=\min_{Z\in\Asc_\rho}\,\max_{\substack{Q\in\Msc_{1,f}(P)\\ E_Q[Z]\ge0}}E_Q[-X],\quad\forall\,X\in L^\infty(P).$$
%where $\Qsc_Z=\{Q\in\Msc_1(P)\mid E_Q[Z]\ge0\}$ for each $Z\in\Asc_\rho$.
Similar to Corollary \ref{cor:convex}, we can see that the proof of Theorem \ref{thm:phrm:rep:P} does not depend on but leads to the classical representation theorem for coherent risk measures of Artzner, Delbaen, Eber and Heath (1999).
\end{remark}

\section{Law-Invariant Risk Measures}\label{sec:LI}

From now on, we assume the probability space $(\Omega,\Fc,P)$ is nonatomic, that is, it supports a random variable with a continuous distribution.
Now we study the class of all risk measures which are law-invariant.

\begin{definition}
A monetary risk measure $\rho: L^\infty(P)\to\Rbb$ is called law-invariant if
$\rho(X)=\rho(Y)$ whenever $X$ and $Y$ have the same distribution under $P$.
\end{definition}

\begin{definition} Given two random variables $X,Y\in L^\infty(P)$.
We say that $X$ first-order stochastically dominates (FSD, in short) $Y$ and write it $X\fsd Y$, if
$$E_P[f(X)]\ge E_P[f(Y)]$$
for all increasing\footnote{Throughout the paper ``increasing" means ``non-decreasing" and ``decreasing" means ``non-increasing."}  functions $f$.
\end{definition}

It is easy to see that a monetary risk measure $\rho: L^\infty(P)\to\Rbb$ is law-invariant if and only if it is FSD-consistent in the following  sense:
\begin{itemize}
\item FSD-Consistency:   $\rho(X)\le\rho(Y)$ whenever $X\fsd Y$.
\end{itemize}

For a random variable $X\in L^\infty(P)$,  its right-continuous probability distribution function is denoted by $F_X$ and its (upper) quantile function
$q_X: [0,1]\to\Rbb$ is given by
\begin{align*}q_X(t)\trieq&\,\inf\{x\in\Rbb\mid F_X(x)>t\}, \quad t\in[0,1),\\
q_X(1)\trieq &\,q_X(1-)=\lim_{t\uparrow 1}q_X(t).
\end{align*}
For more details about quantile functions,
see Appendix A.3 of F\"ollmer and Schied (2016).
One of the most well known law-invariant risk measures is VaR.
The VaR of $X$ at level $t\in[0,1]$ is given by
$$\VaR_t(X)\trieq -q_X(t).$$
%For all $X,Y\in L^\infty(P)$, we have $X$ and $Y$ have the same distribution if and only if $q_X=q_Y$
It is well known that, for any $X,Y\in L^\infty(P)$,
\begin{align}\label{eq:fsd}
X\fsd Y \Leftrightarrow  \VaR_t(X)\le \VaR_t(Y),\ \forall\, t\in[0,1].
\end{align}

Now we introduce some notations.
\begin{itemize}
\item $\Gc$ denotes all nonnegative Borel functions $g$ on $(0,1)$ such that $\int_0^1g(t)dt=1$. That is, $\Gc$ is the set of all probability density functions on $(0,1)$.
\item %We consider the probability space $([0,1],\Bc_{[0,1]}),\ell)$, where
 $\Bsc_{[0,1]}$ denotes all Borel subsets of $[0,1]$.
 \item  $\ell$ denotes the Lebesgue measure on $[0,1]$.
\item $\Msc_{1,f}(\ell)$ denotes all \textit{finitely additive} probability measures on $([0,1],\Bsc_{[0,1]})$ which are absolutely continuous with respect to $\ell$. \end{itemize}

It is well known that $\Msc_{1,f}(\ell)$ is compact under the weak*-topology, which is induced by $L^\infty([0,1],\Bsc_{[0,1]}),\ell)$.

The following theorem characterizes law-invariant monetary risk measures on $L^\infty(P)$.

\begin{theorem}\label{thm:mrm:rep:L}
Assume the probability space $(\Omega,\Fc,P)$ is nonatomic.
For a mapping $\rho: L^\infty(P)\to\Rbb$, the following two assertions are equivalent.
\begin{description}
\item[(a)] $\rho$ is a law-invariant monetary risk measure.
\item[(b)] There exists a family $\{\alpha_\lambda\mid\lambda\in\Lambda\}$ of convex functionals $\alpha_\lambda: \Gc\to(-\infty,\infty]$ such that
\begin{align}\label{eq:mrm:rep:L}
\rho(X)=\min_{\lambda\in \Lambda}\sup_{g\in\Gc}\left(\int_0^1\VaR_t(X)g(t)dt-\alpha_\lambda(g)\right)\quad\mbox{ for }X\in L^\infty(P).
\end{align}
%Here $\VaR(X)$ denotes the function $(0,1)\ni t\mapsto \VaR_t(X)$.
\item[(c)] There exists a family $\{\alpha_\lambda\mid\lambda\in\Lambda\}$ of lower semi-continuous (under the weak*-topology), convex functionals $\alpha_\lambda: \Msc_{1,f}(\ell)\to(-\infty,\infty]$ such that
\begin{align}\label{eq:mrm:rep:L}
\rho(X)=\min_{\lambda\in \Lambda}\,\max_{\mu\in\Msc_{1,f}(\ell)}\left(\int_0^1\VaR_t(X)\mu(dt)-\alpha_\lambda(\mu)\right)\quad\mbox{ for }X\in L^\infty(P).
\end{align}
\end{description}
\end{theorem}

\proof ``(b)$\Rightarrow$(a)" and ``(c)$\Rightarrow$(a)" are obvious.

Now we show ``(a)$\Rightarrow$(b)".
Assume $\rho$ is a law-invariant monetary risk measure. For any
$Z\in\Asc_\rho$, let
$$\Dsc_1(Z)=\{Y\in L^\infty(P)\mid Y\fsd Z\}.$$
By a similar discussion as in the proof of Theorem \ref{thm:mrm:rep:P}, we can see that
each $\rho_{\Dsc_1(Z)}$ is a law-invariant monetary risk measure and
$$\rho(X)=\min_{Z\in\Asc_\rho}\rho_{\Dsc_1(Z)}(X).$$
%Firstly, each $\Dsc_1(Z)$ obviously satisfies conditions (i)--(ii) of Proposition \ref{prop:rm:set} and each $\rho_{\Dsc_1(Z)}$ is a law-invariant monetary risk measure.
%Next, $\Asc_\rho=\bigcup_{Z\in\Asc_\rho}\Dsc_1(Z)$. Then by Lemma \ref{lma:inf},
%$$\rho(X)=\rho_{\Asc_\rho}(X)=\inf_{Z\in\Asc_\rho}\rho_{\Dsc_1(Z)}(X),\quad \forall\, X\in L^\infty(P).$$
%Moreover, for each $X\in L^\infty(P)$, we have $Z_0=X+\rho(X)\in\Asc_\rho$ and
%$$\rho_{\Dsc_1(Z_0)}(X)=\inf\{m\in\Rbb\mid X+m\fsd X+\rho(X)\}=\rho(X),$$
%which implies
%$$\rho(X)=\min_{Z\in\Asc_\rho}\rho_{\Dsc_1(Z)}(X).$$
Moreover, for each $X\in L^\infty(P)$ and $Z\in\Asc_\rho$, we have
\begin{align*}
\rho_{\Dsc_1(Z)}(X)=&\inf\{m\in\Rbb\mid X+m\fsd Z\}\\
=&\inf\{m\in\Rbb\mid \VaR_t(X+m)\le \VaR_t(Z),\ \forall\,t\in(0,1)\}\\
=&\inf\{m\in\Rbb\mid \VaR_t(X)-m\le \VaR_t(Z),\ \forall\,t\in(0,1)\}\\
=&\,\sup_{t\in(0,1)}(\VaR_t(X)-\VaR_t(Z))\\
=&\sup_{g\in\Gc}\int_0^1 (\VaR_t(X)-\VaR_t(Z))g(t)dt,\\
=&\sup_{g\in\Gc}\left(\int_0^1\VaR_t(X)g(t)dt-\alpha_Z(g)\right),
\end{align*}
where $\alpha_Z(g)=\int_0^1\VaR_t(Z)g(t)dt$ for each $g\in\Gc$.
Therefore, $\{\alpha_Z\mid Z\in\Asc_\rho\}$ is a  desired family of convex functionals on $\Gc$.

The proof of ``(a)$\Rightarrow$(c)" is similar to the corresponding part of Theorem \ref{thm:mrm:rep:P}.
\qed

\begin{remark}\label{rmk:mrm:L}
From the proof of Theorem \ref{thm:mrm:rep:L}, we can see that a law-invariant monetary risk measure $\rho: L^\infty(P)\to\Rbb$ has the following representation
\begin{align*}
\rho(X)=&\min_{Z\in\Asc_\rho}\sup_{g\in\Gc}\left(\int_0^1\VaR_t(X)g(t)dt-\int_0^1\VaR_t(Z)g(t)dt\right)\\
=&\min_{Z\in\Asc_\rho}\,\max_{\mu\in\Msc_{1,f}(\ell)}\left(\int_0^1\VaR_t(X)\mu(dt)-\int_0^1\VaR_t(Z)\mu(dt)\right),
\end{align*}
for all $X\in L^\infty(P)$.
\end{remark}

The following theorem characterizes law-invariant, positively homogeneous risk measures on $L^\infty(P)$, whose proof, based on Theorem \ref{thm:mrm:rep:L}, is similar to Theorem
\ref{thm:phrm:rep:P} and hence omitted.

\begin{theorem}\label{thm:phrm:rep:L}
Assume the probability space $(\Omega,\Fc,P)$ is nonatomic.
For a mapping $\rho: L^\infty(P)\to\Rbb$, the following two assertions are equivalent.
\begin{description}
\item[(a)] $\rho$ is a law-invariant, positively homogeneous risk measure.
\item[(b)] There exists a family $\{\Msc_\lambda\mid\lambda\in\Lambda\}$ of nonempty, weak*-compact, convex subsets of $\Msc_{1,f}(\ell)$ such that
$$\rho(X)=\min_{\lambda\in \Lambda}\max_{\mu\in\Msc_\lambda}\int_0^1\VaR_t(X)\mu(dt)\quad\mbox{ for }X\in L^\infty(P).$$
\end{description}
\end{theorem}

\begin{remark}\label{rmk:phrm:L}
Similarly to  Remark \ref{rmk:phrm:rep}, we can see that a law-invariant, positively homogeneous risk measure $\rho: L^\infty(P)\to\Rbb$ has the following representation
$$\rho(X)=\min_{Z\in\Asc_\rho}\,\max_{\substack{\mu\in\Msc_{1,f}(\ell)\\ \int_0^1\VaR_t(Z)\mu(dt)\le0}}\int_0^1\VaR_t(X)\mu(dt),\quad\forall\,X\in L^\infty(P).$$
\end{remark}

\section{SSD-Consistent Risk Measures}\label{sec:ssd}

Now we study SSD-consistent risk measures in this section.

\begin{definition} Given two random variables $X,Y\in L^\infty(P)$.
We say that $X$ second-order stochastically dominates $Y$ and write it $X\ssd Y$, if
$$E_P[f(X)]\ge E_P[f(Y)]$$
for all increasing and concave functions $f$.
\end{definition}

\begin{definition}
A monetary risk measure $\rho: L^\infty(P)\to\Rbb$ is called SSD-consistent if
$\rho(X)\le\rho(Y)$ whenever $X\ssd Y$.
\end{definition}

Mao and Wang (2020) comprehensively investigated SSD-consistent risk measures and provided four equivalent conditions of SSD-consistency. Theorem \ref{thm:idcv} below gives another one:
SSD-consistency is equivalent to the combination of law-invariance, the Fatou property and a kind of weak convexity, called ID-convexity.  As far as we know, the ID-convexity has not yet been introduced in the literature to study risk measures.

\begin{definition} A monetary risk measure $\rho: L^\infty(P)\to\Rbb$ is called ID-convex (identical-distribution convex) if it satisfies
\begin{itemize}
\item ID-Convexity: $\rho(\sum_{i=1}^n\alpha_iX_i)\le\sum_{i=1}^n\alpha_i\rho(X_i)$ whenever $X_1$ $X_2$, \dots, $X_n$ are identically distributed under $P$, each $\alpha_i\ge0$ and $\sum_{i=1}^n\alpha_i=1$.
\end{itemize}
\end{definition}

\begin{definition}
%A monetary risk measure $\rho: L^\infty(P)\to\Rbb$ is continuous from
%above if $$X_n\searrow X\,\Rightarrow\,\rho(X_n)\nearrow\rho(X).$$
We say a monetary risk measure $\rho: L^\infty(P)\to\Rbb$ has the Fatou property if
$$\left.\begin{aligned}
&(X_n)\mbox{ is a bounded sequence in }L^\infty(P)\\
&X_n\rightarrow X\ P\mbox{-a.s.}
\end{aligned}
\right\}\Rightarrow\, \liminf_{n\to\infty}\rho(X_n)\ge \rho(X).$$
\end{definition}

\begin{theorem}\label{thm:idcv}
Assume the probability space $(\Omega,\Fc,P)$ is nonatomic.
For a monetary risk measure $\rho: L^\infty(P)\to\Rbb$, the following two assertions are equivalent.
\begin{description}
\item[(a)] $\rho$ is SSD-consistent.
\item[(b)] $\rho$ is law-invariant and ID-convex and has the Fatou property.
\end{description}
\end{theorem}

\proof
\begin{description}
\item[``(a)$\Rightarrow$(b)":]
Assume $\rho$ is SSD-consistent. The law-invariance of $\rho$ is clear. The Fatou property is from Mao and Wang (2020, Theorem 3.5). It remains to show the ID-convexity. Actually, let $X_1$ $X_2$, \dots, $X_n$ be identically distributed under $P$, each $\alpha_i\ge0$ and $\sum_{i=1}^n\alpha_i=1$.
Obviously, we have
$\sum_{i=1}^n\alpha_iX_i\ssd X_1$ and hence
$$\rho\left(\sum_{i=1}^n\alpha_iX_i\right)\le\rho(X_1)=\sum_{i=1}^n\alpha_i\rho(X_i).$$ Therefore, $\rho$ is ID-convex.
\item[``(b)$\Rightarrow$(a)":] Assume $\rho$ is law-invariant and ID-convex and has the Fatou property. For any $X,Y\in L^\infty(P)$ with $X\ssd Y$, we need to show $\rho(X)\le\rho(Y)$. Actually, by a result of Ryff (1967), see also Carlier and Lachapelle (2011, Lemma 2.3), there exists a sequence $Z_n$ of the form
$Z_n=\sum_{i=1}^{N_n}\alpha_i^nY_i^n$ with $\alpha_i^n\ge0$, $\sum_{i=1}^{N_n}\alpha_i^n=1$ and each $Y_i^n$ has the same distribution of $Y$, such that $Z_n$ converges to $X$ $P$-a.s.. Then by the law-invariance, ID-convexity and the Fatou property of $\rho$, we have
\begin{align*}
\rho(X)\le\liminf_{n\to\infty}\rho(Z_n)\le\liminf_{n\to\infty}\sum_{i=1}^{N_n}\alpha_i^n\rho(Y_i^n)=\rho(Y).
\end{align*}
\qed
\end{description}

The most well known SSD-consistent risk measure is AVaR (average value at
risk\footnote{Also sometimes termed ``expected shortfall" or
``conditional value at risk" in the literature.}).
 The AVaR of $X$ at level $t\in[0,1]$ is given by
\begin{align*}
\AVaR_t(X)\trieq & \,{1\over t}\int_0^t\VaR_s(X)ds,\quad t\in(0,1],\\
\AVaR_0(X)\trieq & \,\VaR_0(X).
\end{align*}
It is well known that, for any $X,Y\in L^\infty(P)$,
\begin{align}\label{eq:sad}
X\ssd Y \Leftrightarrow \AVaR_t(X)\le\AVaR_t(Y),\
\forall\, t\in[0,1].
\end{align}

Let
\begin{align*}
\Wc\trieq\left\{w:[0,1]\to[0,1]\mid 0\le w(0)\le w(1)=1,\ w \mbox{ is
 increasing and right-continuous}\right\}.
\end{align*}
Then $\Wc$ is the set of all probability distribution functions on $[0,1]$. For each probability distribution function $w\in\Wc$, the probability of $\{0\}$ is $w(0)$ and the probability of $\{1\}$ is $1-w(1-)$.
It is well known that $\Wc$ is compact under the weak topology, which is induced by all continuous functions on $[0,1]$.

The results of following two theorems have been reported in Mao and Wang (2020), which are presented and proved here in the manner of the previous two sections.

\begin{theorem}\label{thm:mrm:rep:ssd}
Assume the probability space $(\Omega,\Fc,P)$ is nonatomic.
For a mapping $\rho: L^\infty(P)\to\Rbb$, the following assertions are equivalent.
\begin{description}
\item[(a)] $\rho$ is an SSD-consistent monetary risk measure.
\item[(b)] There exists a family $\{\alpha_\lambda\mid\lambda\in\Lambda\}$ of convex functionals  $\alpha_\lambda: \Gc\to(-\infty,\infty]$ such that
\begin{align}\label{eq:mrm:rep:ssd1}
\rho(X)=\min_{\lambda\in \Lambda}\sup_{g\in\Gc}\left(\int_0^1\AVaR_t(X)g(t)dt-\alpha_\lambda(g)\right)\quad\mbox{ for }X\in L^\infty(P).
\end{align}
\item[(c)] There exists a family $\{\alpha_\lambda\mid\lambda\in\Lambda\}$ of lower semi-continuous (under the weak topology), convex functionals  $\alpha_\lambda: \Wc\to(-\infty,\infty]$ such that
\begin{align}\label{eq:mrm:rep:ssd2}
\rho(X)=\min_{\lambda\in \Lambda}\max_{w\in\Wc}\left(\int_{[0,1]}\AVaR_t(X)dw(t)-\alpha_\lambda(w)\right)\quad\mbox{ for }X\in L^\infty(P).
\end{align}
\item[(d)] There exists a family $\{\rho_\lambda\mid\lambda\in\Lambda\}$ of law-invariant, convex risk measures on $L^\infty(P)$ such that
$$
\rho(X)=\min_{\lambda\in \Lambda}\rho_\lambda(X)\quad\mbox{ for }X\in L^\infty(P).
$$
\item[(e)] For each $X\in L^\infty(P)$,
$$\rho(X)=\inf\{h(X)\mid h\mbox{ is a law-invariant, convex risk measure and }h\ge\rho\}.$$
\end{description}
\end{theorem}

\proof The ``(b)$\Rightarrow$(d)$\Rightarrow$(e)$\Rightarrow$(a)" and ``(c)$\Rightarrow$(d)" parts are obvious.
The proof of ``(a)$\Rightarrow$(b)" and ``(a)$\Rightarrow$(c)" parts is similar to Theorem \ref{thm:mrm:rep:L}.\qed

\begin{theorem}\label{thm:phrm:rep:ssd}
Assume the probability space $(\Omega,\Fc,P)$ is nonatomic.
For a mapping $\rho: L^\infty(P)\to\Rbb$, the following assertions are equivalent.
\begin{description}
\item[(a)] $\rho$ is an SSD-consistent, positively homogeneous risk measure.
\item[(b)] There exists a family $\{\Wc_\lambda\mid\lambda\in\Lambda\}$ of nonempty, weakly compact, convex subsets of $\Wc$ such that
$$\rho(X)=\min_{\lambda\in \Lambda}\max_{w\in\Wc_\lambda}\int_{[0,1]}\AVaR_t(X)dw(t)\quad\mbox{ for }X\in L^\infty(P).$$
\item[(c)] There exists a family $\{\rho_\lambda\mid\lambda\in\Lambda\}$ of law-invariant, coherent risk measures on $L^\infty(P)$ such that
$$
\rho(X)=\min_{\lambda\in \Lambda}\rho_\lambda(X)\quad\mbox{ for }X\in L^\infty(P).
$$
\item[(d)] For each $X\in L^\infty(P)$,
$$\rho(X)=\inf\{h(X)\mid h\mbox{ is a law-invariant, coherent risk measure and }h\ge\rho\}.$$
\end{description}
\end{theorem}

\proof It is similar to Theorem \ref{thm:phrm:rep:P}, based on Theorem \ref{thm:mrm:rep:ssd}.\qed

\begin{remark}\label{rmk:mph:ssd}
Similar to Remarks \ref{rmk:mrm:L} and \ref{rmk:phrm:L}, we have the following two assertions.
\begin{description}
  \item[(a)] An SSD-consistent monetary risk measure $\rho: L^\infty(P)\to\Rbb$ has the following representations
\begin{align*}
\rho(X)=&\min_{Z\in\Asc_\rho}\sup_{g\in\Gc}\left(\int_0^1\AVaR_t(X)g(t)dt-\int_0^1\AVaR_t(Z)g(t)dt\right)\\
=&\min_{Z\in\Asc_\rho}\max_{w\in\Wc}\left(\int_{[0,1]}\AVaR_t(X)dw(t)-\int_{[0,1]}\AVaR_t(Z)dw(t)\right),\quad\forall\,X\in L^\infty(P).
\end{align*}

  \item[(b)] An SSD-consistent, positively homogeneous risk measure $\rho: L^\infty(P)\to\Rbb$ has the following representation
$$\rho(X)=\min_{Z\in\Asc_\rho}\max_{\substack{w\in\Wc\\ \int_{[0,1]}\AVaR_t(Z)dw(t)\le0}}\int_{[0,1]}\AVaR_t(X)dw(t),\quad\forall\,X\in L^\infty(P).$$
\end{description}
\end{remark}

The SSD-consistency obviously implies the law-invariance. Then a question arises: what is the representation in terms of VaR for an SSD-consistent monetary risk measure, when it is regarded as a law-invariant monetary risk measure? The same question arises for SSD-consistent, positively homogeneous risk measures as well. They are positively replied by the next corollary, where the following notations will be used.
\begin{align*}
\Gc^\downarrow\trieq&\{g\in\Gc\mid g\mbox{ is decreasing on }(0,1)\},\\
\Psi\trieq&\{\psi\in\Wc\mid \psi\mbox{ is concave on }[0,1]\}.
\end{align*}
By  F\"ollmer and Schied (2016, Lemma 4.69), the identity
\begin{align}\label{eq:psi}
\begin{cases}
    \psi^\prime(t)=\int_{(t,1]}s^{-1}dw(s),\quad t\in[0,1)  \\
     \psi(0)=w(0)
\end{cases}
\end{align}
defines a bijection
$$J: \Wc\to\Psi,$$ where $\psi^\prime$ denotes the right derivative. Under identity \eqref{eq:psi}, an application of Fubini theorem implies that
\begin{align}\label{eq:int:psi}
\int_{[0,1]}\VaR_t(X)d\psi(t)=\int_{[0,1]}\AVaR_t(X)dw(t),\quad\forall\,X\in L^\infty(P).
\end{align}
Let $\Psi$ be endowed with the topology induced by the bijection $J$.  Then
$\Wc$ and $\Psi$ are homeomorphic and $J$ is a homeomorphism. Therefore,
$\Psi$ is a compact space and, for every $X\in L^\infty(P)$, the functional
$$\psi\mapsto\int_{[0,1]}\VaR_t(X)d\psi(t)$$
is continuous on $\Psi$.

%An application of Fubini theorem implies that, for each $X\in L^\infty(P)$ and $g\in\Gc$,
%\begin{align*}
%\int_0^1\AVaR_t(X)g(t)dt=&\int_0^1\left({1\over t}\int_0^t\VaR_s(X)ds\right)g(t)dt\\
%=&\int_0^1\VaR_s(X)\left(\int_s^1{g(t)\over t}dt\right)ds\\
%=&\int_0^1\VaR_s(X)\tilde g(s)ds,
%\end{align*}
%where $$\tilde g(s)=\int_s^1{g(t)\over t}dt,\quad s\in(0,1).$$
%It is easy to verify that $g\mapsto \tilde g$ is a bijection
%%(one-to-one onto mapping)
%from $\Gc$ to $\Gc^\downarrow$.
%%Moreover, $$\int_0^1\AVaR_t(X)g(t)dt=\int_0^1\VaR_s(X)\tilde g(s)ds.$$
%Moreover, we have $\psi(0)=w(0)$.
%\begin{lemma}\label{lam:psi}
%$\Psi$ is a compact subset of $\Wc$ under the weak topology.
%\end{lemma}

\begin{corollary}\label{cor:mph:ssd}
Assume the probability space $(\Omega,\Fc,P)$ is nonatomic.
For a mapping $\rho: L^\infty(P)\to\Rbb$, we have the following two assertions.
\begin{description}
\item[(a)] $\rho$ is an SSD-consistent monetary risk measure if and only if one of the following conditions holds.
\begin{description}
\item[(i)] There exists a family $\{\alpha_\lambda\mid\lambda\in\Lambda\}$ of convex functionals  $\alpha_\lambda: \Gc^\downarrow\to(-\infty,\infty]$ such that
\begin{align}\label{eq:mrm:rep:Lssd}
\rho(X)=\min_{\lambda\in \Lambda}\sup_{g\in\Gc^\downarrow}\left(\int_0^1\VaR_t(X)g(t)dt-\alpha_\lambda(g)\right)\quad\mbox{ for }X\in L^\infty(P).
\end{align}
\item[(ii)] There exists a family $\{\alpha_\lambda\mid\lambda\in\Lambda\}$ of lower semi-continuous, convex functionals  $\alpha_\lambda: \Psi\to(-\infty,\infty]$ such that
\begin{align*}%\label{eq:mrm:rep:L}
\rho(X)=\min_{\lambda\in \Lambda}\max_{\psi\in\Psi}\left(\int_{[0,1]}\VaR_t(X)d\psi(t)-\alpha_\lambda(\psi)\right)\quad\mbox{ for }X\in L^\infty(P).
\end{align*}
\end{description}
\item[(b)] $\rho$ is an SSD-consistent, positively homogeneous risk measure if and only if there exists a family $\{\Psi_\lambda\mid\lambda\in\Lambda\}$ of nonempty, compact, convex subsets of $\Psi$ such that
$$\rho(X)=\min_{\lambda\in \Lambda}\max_{\psi\in\Psi_\lambda}\int_{[0,1]}\VaR_t(X)d\psi(t)\quad\mbox{ for }X\in L^\infty(P).$$
\end{description}
\end{corollary}

\proof It is an obvious consequence of Theorems \ref{thm:mrm:rep:ssd}  and \ref{thm:phrm:rep:ssd}.
\qed

%based on the following characterization of SSD.
%For any $X,Y\in L^\infty(P)$, $X\ssd Y$ if and only if
%\begin{align*}%\label{eq:sad}
%\int_{[0,1]}\VaR_t(X)d\psi(t)\le \int_{[0,1]}\VaR_t(Y)d\psi(t),\
%\forall\, \psi\in\Psi.
%\end{align*}
%\qed

\begin{remark}\label{rmk:mph:ssd2}
Similar to Remark \ref{rmk:mph:ssd}, an SSD-consistent monetary risk measure $\rho: L^\infty(P)\to\Rbb$ has the following representations
\begin{align*}
\rho(X)=&\min_{Z\in\Asc_\rho}\sup_{g\in\Gc^\downarrow}\left(\int_0^1\VaR_t(X)g(t)dt-\int_0^1\VaR_t(Z))g(t)dt\right)\\
=&\min_{Z\in\Asc_\rho}\max_{\psi\in\Psi}
\left(\int_{[0,1]}\VaR_t(X)d\psi(t)-\int_{[0,1]}\VaR_t(Z)d\psi(t)\right),
\end{align*}
for all $X\in L^\infty(P)$.
% An SSD-consistent, positively homogeneous risk measure $\rho: L^\infty(P)\to\Rbb$ has the following representation
%$$\rho(X)=\min_{Z\in\Asc_\rho}\sup_{\substack{g\in\Gc^\downarrow\\ \int_0^1\VaR_t(Z)g(t)dt\le0}}\left(\int_0^1\VaR_t(X)g(t)dt\right),\quad\forall\,X\in L^\infty(P).$$
\end{remark}

\section{Characterizations in a Unified Form}\label{sec:unif}

We have shown that each monetary risk measure $\rho$ is the lower envelope of a family of convex risk measures; particularly,
$$\rho(X)=\inf\{h(X)\mid h \mbox{ convex and }h\ge\rho\}.$$
When the law-invariance is involved, it is a ``natural" expectation that each law-invariant monetary risk measure is the lower envelope of a family of law-invariant convex risk measures. But each law-invariant convex risk measure is SSD-consistent. Therefore,  the lower envelope of a family of law-invariant convex risk measures must be SSD-consistent as well. As a consequence, the previous ``natural" expectation is not correct. For example,
$$\VaR_t(X)\ne\inf\{h(X)\mid h \mbox{ law-invariant, convex and }h\ge\VaR_t\}.$$
Furthermore, as Theorem \ref{thm:mrm:rep:ssd} shows, each SSD-consistent monetary risk measure $\rho$ is the lower envelope of a family of law-invariant convex risk measures;
particularly,
$$\rho(X)=\inf\{h(X)\mid h \mbox{ law-invariant, convex and }h\ge\rho\}.$$

A natural question arises: what is a ``good" property such that general/law-invariant/SSD-consistent monetary risk measures can be characterized in a unified form as follows?
\begin{itemize}
  \item A monetary risk measure is the lower envelope of a family of ``good" risk measures.
  \item A law-invariant monetary risk measure is the lower envelope of a family of law-invariant, ``good" risk measures.
  \item An SSD-consistent monetary risk measure is the lower envelope of a family of SSD-consistent, ``good" risk measures.
\end{itemize}
The convexity is two strict to unify the characterizations, as we have seen.
It turns out that the ``comonotonic convexity" (CoM-convexity, for short) is the desired ``good" property, which is weaker than the convexity and were introduced by Song and Yan (2006, 2009), as well as by Kou, Peng and Heyde (2013), to investigate risk measures. Some basic definitions and facts about CoM-convex (CoM-coherent) risk measures are summarized in Appendix A.

The following theorem replies to the previous question.

\begin{theorem} For a mapping $\rho: L^\infty(P)\to\Rbb$, we have the following assertions.
\begin{description}
  \item[(a)] $\rho$ is a monetary risk measure if and only if it is the lower envelope of a family of CoM-convex risk measures.
  \item[(b)] Assume $(\Omega,\Fc, P)$ is nonatomic, then $\rho$ is a law-invariant monetary risk measure if and only if it is the lower envelope of a family of law-invariant, CoM-convex risk measures.
  \item[(c)] Assume $(\Omega,\Fc, P)$ is nonatomic, then $\rho$ is an SSD-consistent monetary risk measure if and only if it is the lower envelope of a family of SSD-consistent, CoM-convex risk measures.
\end{description}
\end{theorem}

\proof (a) is an obvious consequence of Theorem \ref{thm:mrm:rep:P}. (c) is a consequence of Theorems \ref{thm:mrm:rep:ssd} and \ref{thm:sy2009}. The ``if" part of (b) is obvious. Now we show the ``only if" part of (b).

Assume $\rho$ is a law-invariant monetary risk measure. Recalling the proof of Theorem \ref{thm:mrm:rep:L}, the set $\Dsc_1(Z)$ is given by $$\Dsc_1(Z)=\{Y\in L^\infty(P)\mid Y\fsd Z\}.$$
It is easy to see that $\Dsc_1(Z)$ is CoM-convex in the following sense:
$$\alpha X+(1-\alpha)Y\in\Dsc_1(Z)\mbox{ whenever }X,Y\in\Dsc_1(Z) \mbox{ are comonotonic and }\alpha\in[0,1].$$
Then it is easy to verify that $\rho_{\Dsc_1(Z)}$ is CoM-convex.
Therefore, $\{\rho_{\Dsc_1(Z)}\mid Z\in\Asc_\rho\}$ is a desired family of law-invariant, CoM-convex risk measures, whose lower envelope is $\rho$.
\qed

\medskip

Similarly, we have the following theorem.
\begin{theorem} For a mapping $\rho: L^\infty(P)\to\Rbb$, we have the following assertions.
\begin{description}
  \item[(a)] $\rho$ is a positively homogeneous risk measure if and only if it is the lower envelope of a family of CoM-coherent risk measures.
  \item[(b)] Assume $(\Omega,\Fc, P)$ is nonatomic, then $\rho$ is a law-invariant, positively homogeneous risk measure if and only if it is the lower envelope of a family of law-invariant, CoM-coherent risk measures.
  \item[(c)] Assume $(\Omega,\Fc, P)$ is nonatomic, then $\rho$ is an SSD-consistent, positively homogeneous risk measure if and only if it is the lower envelope of a family of SSD-consistent, CoM-coherent risk measures.
\end{description}
\end{theorem}

The next two corollaries show that the proofs of Theorem \ref{thm:mrm:rep:L} and Corollary \ref{cor:mph:ssd} lead to another proof of the representation theorems of Song and Yan (2009) for law-invariant/SSD-consistent and CoM-convex/CoM-coherent risk measures.

\begin{corollary}\label{cor:comconvex:L}
Assume $(\Omega,\Fc, P)$ is nonatomic and $\rho: L^\infty(P)\to\Rbb$ is a law-invariant CoM-convex risk measure. Then there exists a  lower-semicontinuous convex functional $\alpha: \Msc_{1,f}(\ell)\to(-\infty,\infty]$ such that
$$\rho(X)=\max_{\mu\in\Msc_{1,f}(\ell)}\left(\int_0^1\VaR_t(X)\mu(dt)-\alpha(\mu)\right),
\quad\mbox{ for }X\in L^\infty(P).
$$
Furthermore, assume $\rho: L^\infty(P)\to\Rbb$ is law-invariant and CoM-coherent. Then there exists a nonempty, weak* compact and convex subset $\Msc\subseteq\Msc_{1,f}(\ell)$ such that
$$\rho(X)=\max_{\mu\in\Msc}\int_0^1\VaR_t(X)\mu(dt),
\quad\mbox{ for }X\in L^\infty(P).
$$
\end{corollary}

\proof Assume $\rho: L^\infty(P)\to\Rbb$ is a law-invariant CoM-convex risk measure.
%Let $\Qc$ denote all increasing, right-continuous, and bounded
%functions defined on $(0,1)$, that is,
%$$\Qc\trieq\left\{q: (0,1)\to\Rbb\,\left|\, q \mbox{ is increasing, right-continuous and bounded}\right.\right\}.$$
Let $\Qc$ denote the set of quantile functions $q_X$ of all random variables $X\in L^\infty(P)$.
Consider a subset $\Qc_\rho$ of $\Qc$ given by
 $$\Qc_\rho=\{q\in\Qc\mid q=q_Z\mbox{ for some }Z\in\Asc_\rho\}.$$
 Then $\Qc_\rho$ is a convex set. Actually, for any $q_1,q_2\in\Qc_\rho$,
 we have $q_i=q_{Z_i}$ for some $Z_i\in\Asc_\rho$, $i=1,2$. Consider random variables $Y_i=q_i(U)$, $i=1,2$, where $U$ is a $(0,1)$-uniformly distributed random variable. Obviously, for each $i$, $Z_i$ and $Y_i$ have the same distribution. Moreover, $Y_1$ and $Y_2$ are comonotonic. Then we have for each given $\alpha\in(0,1)$, $\alpha q_1+(1-\alpha)q_2=q_Y$, where $Y=\alpha Y_1+(1-\alpha)Y_2$. Moreover,
 $$\rho(Y)\le\alpha\rho(Y_1)+(1-\alpha)\rho(Y_2)=\alpha\rho(Z_1)+(1-\alpha)\rho(Z_2)\le0,$$
 which implies that $Y\in\Asc_\rho$ and hence $\alpha q_1+(1-\alpha)q_2\in\Qc_\rho$. Therefore, $\Qc_\rho$ is convex.

The convexity of $\Qc_\rho$ and the weak*-compactness and convexity of $\Msc_{1,f}(\ell)$ allows for an application of the minmax theorem to
the representations in Remark \ref{rmk:mrm:L}, which leads to,
for each $X\in L^\infty(P)$,
\begin{align*}
\rho(X)=&\min_{q\in\Qc_\rho}\,\max_{\mu\in\Msc_{1,f}(\ell)}\left(\int_0^1\VaR_t(X)\mu(dt)+\int_0^1q(t)\mu(dt)\right)\\
=&\max_{\mu\in\Msc_{1,f}(\ell)}\,\inf_{q\in\Qc_\rho}\left(\int_0^1\VaR_t(X)\mu(dt)+\int_0^1q(t)\mu(dt)\right)\\
=&\max_{\mu\in\Msc_{1,f}(\ell)}\left(\int_0^1\VaR_t(X)\mu(dt)-\alpha(\mu)\right),
\end{align*}
where $\alpha(\mu)=\sup_{q\in\Qc_\rho}\left(-\int_0^1q(t)\mu(dt)\right)$ defines the desired lower-semicontinuous convex functional on $\Msc_{1,f}(\ell)$.

Furthetrmore, if $\rho$ is law-invariant and CoM-coherent, then it is easy to see that $\Asc_\rho$ and $\Qc_\rho$ are cones, which implies that
\begin{align*}
\alpha(\mu)=&\sup_{q\in\Qc_\rho}\left(-\int_0^1q(t)\mu(dt)\right)=\sup_{q\in\Qc_\rho,\alpha>0}\left(-\int_0^1\alpha q(t)\mu(dt)\right)\\
=&
\begin{cases}
  0    & \text{ if }\sup_{q\in\Qc_\rho}\int_0^1q(t)\mu(dt)\ge0, \\
   \infty   & \text{otherwise}.
\end{cases}
\end{align*}
Let $$\Msc=\left\{\mu\in\Msc_{1,f}(\ell)\,\left|\,\sup_{q\in\Qc_\rho}\int_0^1q(t)\mu(dt)\ge0\right.\right\}.$$  Then $\Msc$ is the desired subset of $\Msc_{1,f}(\ell)$.
\qed

\begin{corollary}\label{cor:comconvex:ssd}
Assume $(\Omega,\Fc, P)$ is nonatomic and $\rho: L^\infty(P)\to\Rbb$ is an SSD-consistent CoM-convex risk measure. Then there exists a  lower-semicontinuous convex functional $\alpha: \Psi\to(-\infty,\infty]$ such that
$$\rho(X)=\max_{\psi\in\Psi}\left(\int_{[0,1]}\VaR_t(X)\psi(dt)-\alpha(\psi)\right),
\quad\mbox{ for }X\in L^\infty(P).
$$
Furthermore, assume $\rho: L^\infty(P)\to\Rbb$ is SSD-consistent and CoM-coherent. Then there exists a nonempty, compact and convex subset $\Psi_1\subseteq\Psi$ such that
$$\rho(X)=\max_{\psi\in\Psi_1}\int_{[0,1]}\VaR_t(X)\psi(dt),
\quad\mbox{ for }X\in L^\infty(P).
$$
\end{corollary}

\proof   It is similar to Corollary \ref{cor:comconvex:L}, based on Remarks \ref{rmk:mph:ssd2},  and hence omitted.
\qed

%\newpage
\appendix

\section*{Appendix}
\section{Comonotonic Convex Risk Measures}

\begin{definition} Two random variables $X,Y\in L^\infty(P)$ are called comonotonic if there exists some $\Omega_0\in\Fc$ such that $P(\Omega_0)=1$ and
$$(X(\omega)-X(\omega^\prime))(Y(\omega)-Y(\omega^\prime))\ge0\quad\mbox{ for }\omega,\omega^\prime\in\Omega_0.$$
\end{definition}

\begin{definition}
A monetary risk measure $\rho: L^\infty(P)\to\Rbb$ is called comonotonic additive (CoM-additive, in short) if $$\rho(X+Y)=\rho(X)+\rho(Y)$$ whenever $X,Y\in L^\infty(P)$ are comonotonic.
\end{definition}

\begin{definition}
A monetary risk measure $\rho: L^\infty(P)\to\Rbb$ is called comonotonic convex (CoM-convex, in short) if it satisfies
\begin{itemize}
  \item CoM-Convexity: $\rho(\alpha X+(1-\alpha)Y)\le\alpha\rho(X)+(1-\alpha)\rho(Y)$ whenever $X,Y\in L^\infty(P)$ are comonotonic and $\alpha\in[0,1]$.
\end{itemize}
\end{definition}

\begin{definition}
A monetary risk measure $\rho: L^\infty(P)\to\Rbb$ is called a comonotonic coherent (CoM-coherent, in short) if it is CoM-convex and positively homogeneous.
\end{definition}

\begin{definition}  A set function $c: \Fc\to[0,1]$ is called a capacity on $(\Omega,\Fc)$ if $c(\emptyset)=0$, $c(\Omega)=1$, and $c(A)\le c(B)$ whenever $A\subset B$.
A capacity $c$ on $(\Omega,\Fc)$ is called absolutely continuous with respect to $P$ if $c(A)=c(B)$ whenever $P(A\bigtriangleup B)=0$.
\end{definition}

Let $\Csc(P)$ denote all capacities on $(\Omega,\Fc)$ that are absolutely continuous with respect to $P$.
It is well known that a monetary risk measure $\rho: L^\infty(P)\to\Rbb$ is CoM-additive if and only if there exists some capacity $c\in\Csc(P)$  such that $$\rho(X)=\int (-X)\,dc\quad\mbox{ for }X\in L^\infty(P),$$
where $\int (-X)\,dc$ is the Choquet integral of $-X$ with respect to $c$; see, e.g., Schmeidler (1986), and also F\"ollmer and Schied (2016).

Obviously, any convex (coherent) risk measure is CoM-convex (CoM-coherent).
For the following two theorems, which characterize CoM-convex (CoM-coherent) risk measures, see Song and Yan (2006) as well as Kou, Peng and Heyde (2013).

\begin{theorem} A mapping $\rho: L^\infty(P)\to\Rbb$ is a CoM-convex risk measure if and only if there exists some convex functional $\alpha:\Csc(P)\to(-\infty,\infty]$ such that
$$\rho(X)=\sup_{c\in \Csc(P)}\left\{\int (-X)\,dc-\alpha(c)\right\}\quad\mbox{ for }X\in L^\infty(P).$$
\end{theorem}

\begin{theorem} A mapping $\rho: L^\infty(P)\to\Rbb$ is a CoM-coherent risk measure if and only if there exists some nonempty subset $C\subseteq\Csc(P)$ such that
$$\rho(X)=\sup_{c\in C}\left\{\int (-X)\,dc\right\}\quad\mbox{ for }X\in L^\infty(P).$$
\end{theorem}

For the following theorem, which characterizes SSD-consistent, CoM-convex (CoM-coherent) risk measures, see Song and Yan (2009, Theorems 3.2 and 3.6).

\begin{theorem}\label{thm:sy2009} Assume $(\Omega,\Fc,P)$ is nonatomic. For a monetary risk measure $\rho$ on $L^\infty(P)$, we have the following assertions.
\begin{description}
  \item[(a)] $\rho$ is SSD-consistent and CoM-convex if and only if $\rho$ is law-invariant and convex.
  \item[(b)] $\rho$ is SSD-consistent and CoM-coherent if and only if $\rho$ is law-invariant and coherent.
\end{description}
\end{theorem}

\newpage

%\begin{thebibliography}{9}
\section*{References}{\small

\begin{description}

%\item Abdellaoui, M. (2002): ``A Genuine Rank-Dependent Generalization
%of the von Neumann-Morgenstern Expected Utility Theorem,"
%\textit{Econometrica} \textbf{70}, 717--736.

\item Artzner, P., F. Delbaen, J.-M. Eber, and D. Heath (1999): ``Coherent measures of risk," \textit{Mathematical Finance} \textbf{9}, 203--228.

\item Carlier, G. and A. Lachapelle (2011): ``A Numerical Approach for a Class of Risk-Sharing Problems,"
\textit{Journal of Mathematical Economics} \textbf{47}, 1--13.

\item Cerreia-Vioglio, S., F. Maccheroni, M. Marinacci, and L. Montrucchio (2011): ``Risk Measures: Rationality and Diversification,"
\textit{Mathematical Finance} \textbf{21}, 743--774.

\item Delbaen, F. (2002): ``Coherent measures of risk on general probability spaces." In: \textit{Advances in Finance and Stochastics. Essays in Honour of Dieter Sondermann}, Springer-Verlag, 1--37.

%\item Denneberg, D. (1994):
%\textit{nonadditive Measure and Integral}.
% Kluwer Academic Publishers.

%\item Dunford, N., and J. T. Schwartz (1958):
%\textit{Linear Operaters Part I: General Theory}.
%New York: Interscience Publishers.

\item El Karoui, N. and C. Ravanelli (2009): ``Cash subadditive risk measures and interest rate ambiguity," \textit{Mathematical Finance} \textbf{19},  561--590.

\item F\"ollmer, H. and A. Schied (2002): ``Convex measures of risk and trading constraints," \textit{Finance and Stochastics} \textbf{6}, 429--447.

\item F\"ollmer, H. and A. Schied (2016): \textit{Stochastic
Finance: An Introduction in Discrete Time (4th Edition)}. 1st Edition: 2002. Berlin:
Walter de Gruyter.

\item Frittelli, M. and E. Rosazza Gianin (2002): ``Putting order in risk measures," \textit{J. Banking \& Finance} \textbf{26}, 1473--1486.

\item Frittelli, M. and E. Rosazza Gianin (2005): ``Law-invariant convex risk measures," \textit{Advances in Mathematical Economics} \textbf{7}, 33--46.

%\item Huber, P. J. (1981):
%\textit{Robust Statistics}, Section 10.2. New York: Wiley.

\item Jouini, E., W. Schachermayer, and N. Touzi (2006): ``Law invariant risk measures have the Fatou property," \textit{Advances in Mathematical Economics} \textbf{9}, 49--71.

%\item Kahneman, D. and A. Tversky (1979): ``Prospect Theory: An
%Analysis of Decision under Risk," \textit{Econometrica} \textbf{47},
%263--291.

%\item Kou, S. G., X. Peng, and C. C. Heyde (2011):
%``Robust External Risk Measures," In: \textit{Wiley Encyclopedia of
%Operations Research and Management Science} (Edited by J. J.
%Cochran). John Wiley \& Sons.

\item Kou, S. G., X. Peng, and C. C. Heyde (2013): ``External Risk Measures
and Basel Accords," \textit{Mathematics of Operations Research} 38, 393--417.

%\item Kramkov, D. and W. Schachermayer (1999): ``The Asymptotic Elasticity of Utility Functions and
%Optimal Investment in Incomplete Markets," \textit{Ann. Appl. Prob.}
%\textbf{9}, 904--950.

\item Kusuoka, S. (2001): ``On law invariant coherent risk measures," \textit{Advances in Mathematical Economics} \textbf{3}, 83--95.

\item Mao, T., and R. Wang (2020): ``Risk aversion in regulatory capital principles," \textit{SIAM Journal on Financial Mathematics} 11, 169--200.

\item Mertens, J.-F., S. Sorin, and S. Zamir (2015): \textit{Repeated Games}. New York: Cambridge University Press.

%\item Prelec, D. (1998): ``The Probability Weighting Function,"  \textit{Econometrica} \textbf{66},
%497--527.

%\item Quiggin, J. (1982):
%``A Theory of Anticipated Utility,"
%\textit{Journal of Economic Behavior and Organization} 3, 323--343.
%
%\item Quiggin, J. (1993):
%\textit{Generalized Expected Utility Theory: The Rank-Dependent Model}.
%Boston: Kluwer Academic Publishers.

\item Ryff, J. V. (1967): ``Extreme Points of Some Convex Subsets of $L^1$," \textit{Proceedings of the American Mathematical Society} \textbf{116}, 1026--1034.

%\item Schachermayer,  W. (2001): ``Optimal Investment in Incomplete Markets When Wealth May Become Negative," \textit{Ann. Appl. Prob.}
%\textbf{11}, 694--734.

\item Schmeidler, D. (1986):
``Integral Representation without Additivity," \textit{Proceedings of
the American Mathematical Society} 97, 255--261.

%\item Schmeidler, D. (1989):
%``Subject Probability and Expected Utility without Additivity,"
%\textit{Econometrica} 57, 571--587.

\item Song, Y., and J.-A. Yan (2006):
``The representations of two types of functions on
$L^\infty(\Omega,\Fc)$ and $L^\infty(\Omega,\Fc,\Pbb)$,"
\textit{Science in China Series A: Mathematics} 49, 1376--1382.

\item Song, Y., and J.-A. Yan (2009):
``Risk Measures with Comonotonic Subadditivity or Convexity and
Respecting Stochastic Orders," \textit{Insurance: Mathematics and
Economic} 45, 459--465.

%\item Starmer, C. (2000): ``Developments in Non-Expected Utility Theory: The Hunt for a Descriptive
%Theory of Choice under Risk," \textit{Journal of Economic
%Literature} \textbf{38}, 332--382.

%\item Tversky, A. and D. Kahneman (1992): ``Advances in Prospect
%Theory: Cumulative Representation of Uncertainty," \textit{J. Risk
%Uncertainty} \textbf{5}, 297--323.

%\item von Neumann, J., and O. Morgenstern (1947):
%\textit{Theory of Games and Economic Behavior}, 2nd ed.
%Princeton: Princeton University Press.

%\item Wakker, P. (2010): \textit{Prospect Theory}. Cambridge, UK: Cambridge
%University Press.

%\item Xia, J. and X. Y. Zhou (2016): ``Arrow-Debreu Equilibria for Rank-Dependent
%Utilities," \textit{Mathematical Finance} \textbf{26}, 558--588.
%
%\item Xu, Z. Q. (2016): ``A Note on the Quantile Formulation," \textit{Mathematical Finance} \textbf{26}, 589--601.

%\item Yaari, M. E. (1987): ``The Dual Theory of Choice under Risk,"
%\textit{Econometrica} \textbf{55}, 95--115.

\item Xia, J. (2020): ``Decision Making under Uncertainty: A Game of Two Selves," working paper.

\end{description}
%\end{thebibliography}

\end{document}